\documentclass[superscriptaddress,preprint,longbibliography,amsmath,amssymb,aps,pra,floatfix]{revtex4-2}

\usepackage{graphicx,setspace}
\usepackage{braket}
\usepackage{color,soul}
\usepackage{siunitx}
\usepackage{multirow}
\usepackage{booktabs}

\begin{document}

\title{Blueprint for Diamond Magnetometry: Unraveling Quantum Dephasing of Nitrogen-Vacancy Center Ensembles in Diamond}%

\author{Jixing Zhang}%
\email[]{jixing.zhang@pi3.uni-stuttgart.de}
\affiliation{3rd Institute of Physics, University of Stuttgart, Allmandring 13, Stuttgart 70569, Germany}
	
\author{Cheuk Kit Cheung}%
\affiliation{3rd Institute of Physics, University of Stuttgart, Allmandring 13, Stuttgart 70569, Germany}

\author{Michael K\"ubler}%
\affiliation{3rd Institute of Physics, University of Stuttgart, Allmandring 13, Stuttgart 70569, Germany}
	
\author{Magnus Benke}%
\affiliation{3rd Institute of Physics, University of Stuttgart, Allmandring 13, Stuttgart 70569, Germany}
	
\author{Mathis Brossaud}%
\affiliation{Université Paris-Saclay, ENS Paris-Saclay, DER de Physique, 91190, Gif-sur-Yvette, France.}
	
\author{Andrej Denisenko}%
\affiliation{3rd Institute of Physics, University of Stuttgart, Allmandring 13, Stuttgart 70569, Germany}
	
\author{Ruoming Peng}%
\affiliation{3rd Institute of Physics, University of Stuttgart, Allmandring 13, Stuttgart 70569, Germany}
	
\author{Jens Anders}%
\affiliation{Institute of Smart Sensors, University of Stuttgart, Paffenwaldring 47, Stuttgart 70569, Germany}
	
\author{Emilio Corcione}%
\affiliation{Institute for System Dynamics, University of Stuttgart, Waldburgstr. 17/19, Stuttgart 70563, Germany}
	
\author{Cristina Tar\'{i}n Sauer}%
\affiliation{Institute for System Dynamics, University of Stuttgart, Waldburgstr. 17/19, Stuttgart 70563, Germany}

\author{Andrew M. Edmonds }%
\affiliation{Element Six Global Innovation Centre, Fermi Avenue, Harwell Oxford, Didcot, Oxfordshire OX11 0QR, United Kingdom}

\author{Matthew Markham}%
\affiliation{Element Six Global Innovation Centre, Fermi Avenue, Harwell Oxford, Didcot, Oxfordshire OX11 0QR, United Kingdom}

\author{Kazuo Nakamura}%
\affiliation{Hydrogen $\&$ Carbon Management Technology Strategy Dept., Tokyo Gas Co., Ltd.,  Yokohama 230-0045, Japan}
	
\author{Hitoshi Sumiya}%
\affiliation{Advanced Materials Laboratory, Sumitomo Electric Industries, Ltd., Itami 664-0016, Japan}
	
\author{Shinobu Onoda}%
\affiliation{Quantum materials and applications research center, National Institutes for Quantum Science and Technology, Takasaki, 370-1292 Japan}
	
\author{Junichi Isoya}%
\affiliation{Faculty of Pure and Applied Sciences, University of Tsukuba, Tsukuba 305-8573, Japan}
	
\author{Chen Zhang}%
\email[]{chen.zhang@cquterd.cn}
\affiliation{Beijing Quantum Technology Research and Development Center, Cuiyunxi Rd. 1, Beijing 100074, China}
	
\author{Jörg Wrachtrup}%
\affiliation{3rd Institute of Physics, University of Stuttgart, Allmandring 13, Stuttgart 70569, Germany}
\affiliation{Max Planck Institute for Solid State Research, Stuttgart, Germany}

\begin{abstract}
Diamonds with nitrogen-vacancy (NV) center ensembles are one of the most promising solid-state quantum platforms for various sensing applications.
The combination of a long spin dephasing time ($T_2^*$) and a high NV center concentration is crucial for pushing the sensitivity limits.
In this work, we propose a systematic measurement approach to quantify the electron spin dephasing in NV center ensembles and analyze the contributions of various sources to the dephasing time, including NV-NV interactions, strain and electric field distributions, $^{13}$C nuclear spins, and P1 electron spins.
Our method is validated using a series of high-performance diamond samples, providing a comprehensive understanding of dephasing mechanisms and revealing correlations between NV concentration and different dephasing sources. 
Based on these insights, we further evaluate and propose strategies to improve the achievable sensitivity limits for DC magnetic field measurements.
\end{abstract}

\maketitle


\section{\label{Sec I}Introduction}
Over the past two decades, defects in solid-state materials have been extensively studied as spin systems for quantum technologies\cite{RN1, RN3,61NatureReveiwsMater2018Joerg}.
Among all the known defects, the negatively charged nitrogen-vacancy (NV) center in diamond is one of the most promising platforms for various quantum applications \cite{RN52, RN35, RN40,38nature2022Taminiau}.
Compared to quantum sensors that have demonstrated ultra-high sensitivity in applications such as magnetoencephalography (MEG)\cite{RN5}, magnetometry based on NV centers in diamond is superior in applications that require high spatial resolution and small sensor volumes.
For example, techniques using a single NV center near the diamond surface or in a diamond tip can probe nanoscale magnetization to recognize the structure of protein molecules and 2D materials\cite{RN7, RN44,du2024single}.
Beyond the nanometer scale, diamond magnetometry spatial resolution in the range of micrometers to millimeters with high sensitivity \cite{RN10} is considered a potential technique for applications in magnetic imaging \cite{37nanophotonics2019walsworth}, for example, searching for axions \cite{RN9}. 

The moderate sensitivity of NV ensemble magnetometry remains a major obstacle for practical applications \cite{RN41}.
A sensitivity approaching or better than 1 \unit{pT/\sqrt{Hz}} requires a careful trade-off between defect concentration and the dephasing time $T_2^*$ of NV center ensembles \cite{22RevModPhys2020Walsworth}.
The typical dephasing rate $\Gamma_2^*=1/T_2^*$ of NV center ensembles ranges from tens of kHz up to several MHz, while the defect density ranges from $10^{16}\ \unit{cm^{-3}}$ to $10^{18}\ \unit{cm^{-3}}$ \cite{66MQT2021walsworth}.
For reference, a \unit{fT/\sqrt{Hz}} level atomic magnetometer has atomic vapor with a density of approximately ${10^{14}\ \unit{cm^{-3}}}$ while the decoherence rate is a few Hz \cite{RN12}.
Given the typically high defect density in diamonds, understanding the dephasing mechanisms of NV center ensembles is crucial for extending the coherence time in applications requiring both high sensitivity and spatial resolution. Such understanding will also enable optimization of diamond growth and defect treatment processes, thereby advancing magnetometer performances \cite{RN65, 66MQT2021walsworth}.
Furthermore, the NV center ensemble as a quantum many-body system offers a suitable framework for probing the quantum thermodynamics and simulation, in which unraveling the dephasing sources is fundamentally important\cite{23PhysRevLett2018Lukin, RN26, RN53}.

Previous studies have focused on identifying the primary dephasing sources in specific diamond samples.
For samples containing abundant paramagnetic spin defects, experimental studies combined with theoretical modeling have investigated the spin bath-induced decoherence, enabling the determination of paramagnetic spin concentrations \cite{4PhysRevB2022Walsworth, 28PhysRevB2008Hanson, Wang2023PhdThesis, 1PhysRevMaterials2019Jayich, 3PhysRevB2021Budker}.
Additionally, the linewidth broadening effects induced by the electric fields and strains in diamonds are investigated by both optically detected magnetic resonance (ODMR) and interferometry \cite{46PhysRevLett2018Yao, 32PhysRevApplied2022Walsworth}.
Studies have also addressed NV-NV interactions in diamonds with high NV center concentrations\cite{24NanoLetters2019Jelezko,23PhysRevLett2018Lukin}.
As the dephasing rate of NV center ensembles is influenced by multiple factors and varies significantly across different samples, resulting in diverse quantum sensing performances, we have developed a systematic characterization methodology to unravel these dephasing sources. However, isolating and quantifying individual contributions from the numerous dephasing mechanisms remains a significant technical challenge.

In this work, we present a comprehensive study using multiple experimental techniques to identify dephasing mechanisms of NV ensembles in a series of high-quality diamonds under ambient conditions. 
Our methodology enables systematic characterization of key diamond properties, including electric field and strain distributions, nuclear and electron spin bath properties, and NV center densities. We applied this methodology to characterize diamond samples with varying NV concentrations. 
This enabled us to verify the relationships among electric field strength, Hahn Echo-measured decoherence rates, and NV concentration, while evaluating existing models that describe how different sources contribute to the total dephasing rate. 
Based on these findings, we investigate the fundamental DC sensitivity limits of NV ensemble magnetometry and propose practical strategies to extend the coherence time $T_2^*$, providing a pathway toward enhanced magnetometer performance.

\section{\label{sec II}Experiments}
\subsection{\label{sec IIA}A brief overview on dephasing mechanisms}
We employ a simplified ground-state Hamiltonian for $N$ NV centers, expressed as
\begin{equation}
	{\cal H}_{NV}= \sum_{i=1}^{N} (D_{gs}+\delta^{i}) {S_z^i}^2 + \gamma_e\vec{B}_0\cdot \vec{S^i} + {A^i}{S^i_z} + \sum_{i \neq j} \Vec{S^i}\cdot V_{dd} \cdot \Vec{S^j}.
	\label{Eq_NVHamil}
\end{equation}
Here, $D_{gs}=2.87\ \unit{GHz}$ denotes the zero-field splitting term,  $\delta^{i}$ accounts for the combined effects of local electric fields and strain at each NV center, $\gamma_e$ is the electron gyromagnetic ratio, $\vec{B}_0$ is the magnetic field, $A^i$ denotes the average coupling strength arising from the spin bath (including both electron and nuclear spins), $V_{dd}$ is the dipole-dipole interaction matrix between two NV centers.
In the experiments, a magnetic field $\vec{B}_0 \approx 10\ \unit{G}$ is applied along one NV axis (referred to as the on-axis orientation), while its projections on the other three axes (referred to as off-axis orientations) are identical.
As the bias field dominates over the perpendicular components, the latter are neglected in the equation.

Except for the second term (the Zeeman term), all other terms contribute to the dephasing of NV center ensembles.
These dephasing-related terms can be classified into two groups.
The first group, referred to as inhomogeneous broadening, includes $\delta^i$ and $A^i$, which induce disorder among NV centers.
Specifically, $\delta^i$ arises from inhomogeneous strain distributions and in situ electric field noise, while $A^i$ originates from the nuclear spin bath (associated with $^{13}$C) and the electron spin bath due to paramagnetic defects, such as P1 centers.
The second group, referred to as homogeneous broadening, is dominated by NV-NV interactions, as represented by the last term in Eq.~\ref{Eq_NVHamil}.
	
Assuming the dephasing sources are independent and result in an exponential decay, the dephasing rate of NV center ensembles can be expressed as a summation model~\cite{22RevModPhys2020Walsworth}:
\begin{equation}
	\Gamma_2^* \approx  {\Gamma_{\text{strain}}} + {\Gamma_{\text{elec}}} +{\Gamma_{{13}\text{C}}} + \Gamma_{\text{P1}} + {\Gamma_{\text{NV-NV}}} + {\Gamma_{\text{other}}}+{\Gamma_{1}}.
	\label{eq_summation}
\end{equation}
Here, each term on the right-hand side represents the dephasing rate contributed by the corresponding mechanisms described above.
The term ${\Gamma_\text{other}}$ encompasses all dephasing sources not included in the primary mechanisms.
The longitudinal relaxation time, $T_1=1/\Gamma_1$, sets the upper limit for $T_2^*$. At room temperature, $T_1$ is approximately 6 ms and can be safely neglected in this context~\cite{55PhysRevResearch2021Kolkowitz}.
In the following subsections, we describe the methods employed to investigate each dephasing source individually.
A specific sample is used as an example to demonstrate precise measurements of the dephasing rates in the subsequent sections.

\subsection{\label{sec IIB}Strain and electric field noise}
	
\begin{figure}[b]
    \centering
    \includegraphics{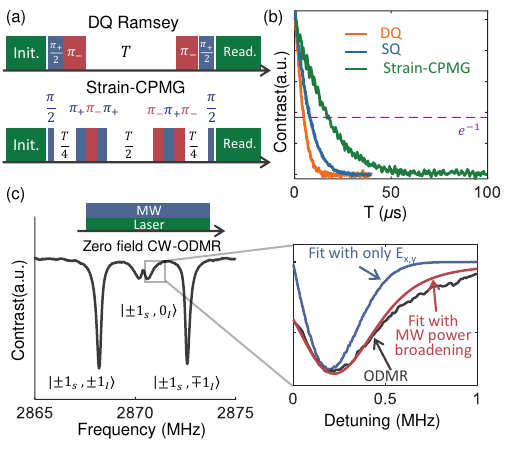}
    \caption{(a) Double quantum Ramsey sequence (upper) and strain-CPMG sequence (lower). (b) Dephasing rate measurements from the example sample. (c) Zero-field ODMR spectrum, with an inset of the enlarged hyperfine line corresponding to the $\ket{m_I = 0}$ state. The line is fitted by considering both strain distribution and microwave power broadening.}
    \label{FigStrain}
\end{figure}

The strain distribution in diamonds is inherently inhomogeneous due to imperfections in material fabrication.
Similarly, the in-situ electric field experienced by NV centers is inhomogeneous, caused by the random spatial distribution of charged defects around the spins.
Both strain and electric fields affect the potential acting on NV electrons, thereby modifying the zero-field splitting term in the ground state.
Consequently, both effects manifest as quadratic terms in the Hamiltonian, represented by $\delta^i$ in Eq.~\ref{Eq_NVHamil} (details provided in Appendix \ref{Appendix A}).
Given that the magnetic dipole effect is linear, experimental sequences can be designed to selectively probe quadratic or linear terms by applying two microwave fields resonant with the $\ket{0}\rightarrow\ket{+1}$ and $\ket{0}\rightarrow\ket{-1}$ transitions, respectively.
To measure these effects, two experimental protocols are used: the double quantum (DQ) Ramsey sequence and the strain-CPMG sequence~\cite{9PhysRevX20185Walsworth, 32PhysRevApplied2022Walsworth}, as illustrated in Figure \ref{FigStrain}(a).

While the DQ Ramsey sequence is insensitive to the $S_z^2$ term in ${\cal{H}}_{NV}$, it is twice as sensitive to the $S_z$ term compared to the conventional Ramsey sequence.
Therefore, $\Gamma_{\text{DQ}}/2$ is used to isolate the dephasing contribution of the dipolar noise.
In contrast, the strain-CPMG sequence facilitates a coherent state swap between $\ket{+1}$ and $\ket{-1}$.
Consequently, the dephasing observed in the strain-CPMG measurement is primarily determined by the $S_z^2$ term in ${\cal{H}}_{NV}$, while remaining insensitive to the $S_z$ term.
Moreover, $\Gamma_{\text{strain-cpmg}}$ incorporates contributions from $\Gamma_2$, which can be estimated through Hahn-echo measurements.
Summarizing these relationships, we obtain:
\begin{equation}\label{eq_strain1}
	{\Gamma_{S_z^2}}={\Gamma_{\text{strain}}}+{\Gamma_{\text{elec}}} = {\Gamma_{\text{strain-cpmg}}-\Gamma_2} \approx {\Gamma_{2}^*} - {\Gamma_{\text{DQ}}/2},
\end{equation}
where $\Gamma_{S_z^2}$ represents the dephasing rate induced by the $S_z^2$ term in ${\cal H}_{NV}$.

In Figure \ref{FigStrain}(b), measurements on the example sample yield $\Gamma_2^* = 137\ \unit{kHz}$, $\Gamma_{\text{DQ}} = 216\ \unit{kHz}$, $\Gamma_2 = 10\ \unit{kHz}$, and $\Gamma_{\text{strain-CPMG}} = 60\ \unit{kHz}$.
From these results, it is evident that the total dephasing rate $\Gamma_2^*$ is only an approximate summation of the dephasing contributions from the $S_z$ and $S_z^2$ terms, rather than precisely satisfying Eq.~\ref{eq_strain1}. 
This discrepancy can be attributed to simplifications in the summation dephasing model. The validity of this model is further examined in the discussion section with more samples.

To further distinguish the dephasing rates contributed by strain and electric fields, it is essential to account for the transverse components induced by strain and electric fields, which are neglected in Eq.~\ref{Eq_NVHamil}.
The parallel and perpendicular terms of the quadratic spin terms in the NV Hamiltonian, caused by isotropic strain and electric field noise, are not equivalent.
Notably, the transverse electric dipole moment of the electric field results in a dip in the zero-field ODMR spectrum~\cite{46PhysRevLett2018Yao}, as shown in Figure \ref{FigStrain}(c) (details in Appendix \ref{Appendix A}).
Due to the hyperfine splitting introduced by the $^{14}N$ nuclear spin ($I = 1$), the effect of the perpendicular term is suppressed for the resonant peaks associated with the $\ket{m_I = \pm 1}$ states.
Only the peak corresponding to the $\ket{m_I = 0}$ state exhibits a dip. The fitted linewidth of this dip, measured from the zero-field ODMR spectrum of the example sample, is approximately $\nu_{\text{zf-ODMR}} = 220\ \unit{kHz}$.
The dephasing rate induced by the electric field is determined as $\Gamma_{\text{elec}} = \nu_{\text{zf-ODMR}} \cdot d^{\parallel}/d^{\perp} = 4.5\ \unit{kHz}$, where $d^{\parallel} = 0.35\ \unit{Hz \cdot cm/V}$ and $d^{\perp} = 17\ \unit{Hz \cdot cm/V}$ are the longitudinal and transverse parameters of the electric dipole moments.
The dephasing rate introduced by the strain distribution is then estimated as $\Gamma_\text{strain}={\Gamma_{S_z^2}}-\Gamma_\text{elec}$.

\subsection{\label{sec IIC}Dipolar noise from the spin bath}

Dipolar noise from the spin bath significantly contributes to the dephasing of NV center ensembles.
Randomly distributed electron and nuclear spins around NV centers create distinct local magnetic fields at each NV center.
Consequently, the spin bath induces a global exponential decay, with a rate proportional to the concentration of bath spins, due to variations in the energy splitting of individual NV centers.
The contribution of the spin bath to the dephasing rates can be estimated using mean-field theory (details in Appendix \ref{Appendix B})~\cite{RN27}.
Accordingly, the dephasing rate induced by the spins is linearly proportional to the spin density, allowing the estimation of spin concentrations from measured dephasing rates, and vice versa.

The contribution of the $^{13}\text{C}$ nuclear spin bath to the dephasing rate is first estimated.
Using the mean-field model (Appendix \ref{Appendix B}), the contribution of $^{13}\text{C}$ (gyromagnetic ratio $\gamma_C = 10.7\ \unit{MHz/T}$) per unit density is derived as ${\Gamma_{\text{13C}}} = 0.1\ \unit{kHz/ppm}.$
The example sample is isotope purified to $0.03\%\ ^{13}\text{C}$, corresponding to a concentration of 300 ppm. 
Hence, $\Gamma_{\text{13C}}=30\ \unit{kHz}$.
For diamonds with a natural abundance of $^{13}\text{C}$ (1.1\%), the $^{13}\text{C}$-induced dephasing rate is approximately $1.1\ \unit{MHz}.$ Consequently, the $T_2^*$ of non-isotope-purified diamonds is typically shorter than $1\ \unit{\mu s}$, which limits the sensitivity of NV ensemble magnetometry.
Besides $^{13}\text{C}$, contributions from other nuclear spins (mostly $^{14}N$ nuclear spins associated with defects such as P1 centers) can be neglected due to their low concentration.

\begin{figure}[t]
    \centering
    \includegraphics{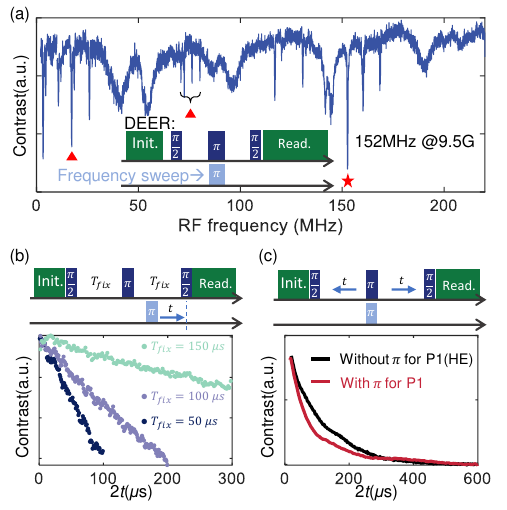}
    \caption{(a) Measured DEER spectrum at a 9.5 Gauss magnetic field. The peaks marked by triangles are related to the dark spin transitions which do not belong to P1 center, while other peaks are P1 center related. The peak marked by star mark is used in measurement in (b) and (c). (b) Pulse-sweep DEER sequence and the corresponding measured signals with the example sample. (c) Duration-sweep DEER sequence and the corresponding measured signals from the same sample.}
    \label{figP1}
\end{figure}

With $^{12}\text{C}$ enriched diamonds, the primary spin bath source shifts to the electron spin bath, predominantly arising from unpaired electrons localized at substituted nitrogen atoms (P1 centers) in the lattice.
Double electron-electron resonance (DEER) measurements are used to evaluate the influence of electron spins (P1 centers \cite{1PhysRevMaterials2019Jayich,24NanoLetters2019Jelezko,11PhysStatusSolidiRRL2022Akimov,RN013} and other dark spins \cite{RN17}) on dephasing.
As shown in Figure \ref{figP1}(a), the DEER spectrum reveals the resonance peaks corresponding to all bath spins within the bandwidth.
In the example, a 9.5 Gauss magnetic field is applied.
The peaks show a complex transition frequencies distribution influenced by diverse energy levels.
Details on the derivation of transition lines are provided in Appendix \ref{Appendix C1}.
An RF pulse, tuned to the frequency of a specific resonance peak in the DEER spectrum, is applied during the Hahn-echo measurement, as shown in Figures \ref{figP1}(b) and \ref{figP1}(c), to characterize the dephasing caused by bath spins and estimate their concentrations.
The resonance peak marked with a star in Figure \ref{figP1}(a) is selected for this example sample to demonstrate the measurement process.
	
In the pulse-sweep DEER measurement illustrated in Figure \ref{figP1}(b), the signal decay follows the expression $\exp(-k_\text{HE}T_\text{fix})\exp(-k_\text{nm}t)$, where $k_{nm}$ represents the decay rate corresponding to the transition line of P1 from the n\textit{th} state to the m\textit{th} state.
However, experimental results exhibit linearly decaying signals rather than exponential decay.
This behavior arises from the relatively low concentration of P1 centers, which results in $\Gamma_{P1} < \Gamma_2$ and restricts the interval $T_{\text{fix}}$ of the sequence to the $T_2$ limit in the pulse-sweep DEER.
As a result, only the initial P1-induced decay curve is observable in the displayed results, appearing as an approximate linear decay.
Consequently, the pulse-sweep DEER measurement is unsuitable for identifying the electron spin bath decay in high-quality diamond samples with low concentrations of paramagnetic defects.

To address the limitations of the pulse-sweep DEER measurement, we propose using a duration-sweep DEER measurement, as illustrated in Figure \ref{figP1}(c).
The signal decay follows $\exp(-(\Gamma_\text{P1, DEER} + \Gamma_{2})t)$, enabling the sequence to determine the decay contribution of P1 centers, even at low concentrations.
For the example sample, $\Gamma_\text{P1, DEER} + \Gamma_\text{2} = 12.9\ \unit{kHz}$ is measured using the sequence with the DEER RF pulse, while $\Gamma_\text{2} = 9.5\ \unit{kHz}$ is obtained using the sequence without it, yielding $\Gamma_\text{P1, DEER} = 3.4\ \unit{kHz}$ as the estimated dephasing rate of $\Gamma_{\text{P1}}$.

Typically, the mix of eigenstates of P1 centers is suppressed at a high magnetic field (\textgreater 100 Gauss) so that the DEER measurement results follow a general mean-field theory \cite{3PhysRevB2021Budker}.
However, in most practical applications, including this work, the magnetic fields are relatively weak.
Thus, the model is adjusted to account for the altered eigenstates of P1 centers under weak magnetic fields (see Appendix \ref{Appendix C1}), and the relationship between $\Gamma_{\text{P1}}$ and $\Gamma_\text{P1, DEER}$ is recalibrated, yielding $\Gamma_{\text{P1}} = 6\Gamma_\text{P1, DEER} = 20.4\ \unit{kHz}.$
The dependency of $\Gamma_{\text{P1}}$ on the P1 center concentration is also remodeled based on the week magnetic field (see Appendix \ref{Appendix C1}).
Moreover, additional peaks not corresponding to P1 transitions are observed in the DEER spectrum, as illustrated in Figure \ref{figP1}(a).
The lines possibly originate from dark spins associated with various defects\cite{RN62,12PhysRevApplied2022SenYang}.
In the example sample, the dephasing contribution from these spins is minimal.
The DEER results of P1 and other dark spins' energy levels, is also a critical prerequisite for spin bath driving, as shown in Appendix \ref{Appendix C2}~\cite{9PhysRevX20185Walsworth,5NatMaterials2014Atature,6SciRep2012Hanson,16npjQuantumInformation2022Jayich}.

\subsection{\label{sec IID}Dipole-dipole interactions among NV centers}
	
\begin{figure}[t]
    \centering
    \includegraphics{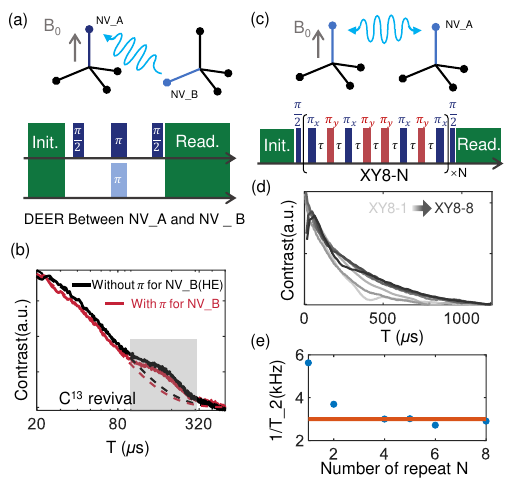}
    \caption{(a) Interaction between NV centers of different groups and the sequence for the corresponding DEER measurement. (b)  Decay rate data (red) obtained using the DEER sequence, with Hahn-Echo results (black) shown for comparison. (c) Interaction between NV centers of the same group and the XY8-N sequence for the $\Gamma_{\text{NV-NV}}$ measurement. (d) Dephasing measurement using XY8-N sequences with different repetition numbers $N$. (e) Dependence of the dephasing rate on the repetition number $N$."}
    \label{FigNVNV}
\end{figure}
	
In most cases, the contribution of dipole-dipole interactions among NV centers to dephasing is relatively small compared to other dephasing sources.
Nevertheless, accurately measuring this contribution is still important, especially for extending $T_2^*$ in samples with high NV center concentrations and low concentrations of other defects~\cite{24NanoLetters2019Jelezko}.
Furthermore, NV center concentrations can be precisely estimated by accurately measuring NV-NV interactions.
Together, these factors—extended $T_2^*$ and precise NV concentration estimation—enable further optimization of the sensitivity of diamond magnetometry.

Diamond samples typically contain NV centers aligned along all four crystal orientations.
We thus need to consider interactions both between NV centers along different orientations (off-axis) and interactions of NV centers along the same axis (on-axis).
Similar to the experiments in Section \ref{sec IIC}, a DEER sequence—with the P1-driving RF pulse replaced by a microwave pulse—can be used to investigate the strength of off-axis NV interactions, as illustrated in Figure \ref{FigNVNV}(a).
For the example sample, the measured additional decay rate induced by off-axis NV interactions is $\Gamma_\text{NV-off} = 1.2\ \unit{kHz}$ (Figure \ref{FigNVNV}(b)), corresponding to an NV$^-$ concentration of $0.12\ \unit{ppm}$ (see Appendix \ref{Appendix D1}).
However, certain limitations need to be addressed in practical applications, as outlined below.
Firstly, imperfections in the microwave $\pi$ pulse can reduce dephasing in the DEER measurement, leading to an underestimation of NV center concentrations~\cite{1PhysRevMaterials2019Jayich, 3PhysRevB2021Budker}.
Secondly, this method becomes invalid if the NV center ensemble in the sample exhibits a preferential orientation~\cite{67APL2014Wrachtrup}.
Moreover, some samples deviate from the exponential decay predicted by mean-field theory (e.g., sample C3 discussed in Section \ref{Sec III}).

To address this limitation, an alternative method utilizing pulse sequences for Hamiltonian engineering is proposed. 
These sequences are designed to ensure that the effective Hamiltonian contains exclusively the NV-NV interaction term, enabling the determination of NV-NV interaction strength by measuring the decay rate corresponding to the specific effective Hamiltonian~\cite{RN63, RN64}.
In theory, dynamic decoupling sequences consisting solely of $\pi$ pulses can fulfill these requirements and two approaches are available for their implementation. 
In the first approach, the $\pi$ pulse interval time \(\tau\) is fixed and kept sufficiently short to satisfy the approximation conditions of Average Hamiltonian Theory (AHT), while the repetition number \(N\) is gradually increased to measure the \(T_2\) decay time, retaining only NV-NV interactions. 
Alternatively, in the second approach, \(\tau\) is incrementally increased with a fixed \(N\) to determine the interaction strength, provided that \(N\) is sufficiently large and \(\tau\) remains short enough to meet the AHT approximation conditions.
In our experiments, we adopted the second approach as it aligns better with commonly used practices in the field.

To ensure that the effective Hamiltonian contains only the NV-NV interaction term, it is crucial to choose a sequence capable of canceling multiple noise components, including those introduced by DC disorder, AC disorder, finite pulse durations, pulse errors, and other sources.
A series of simulations were conducted and confirmed that the XY8 and XY16 sequences are sufficiently robust against the above noise(see Appendix \ref{Appendix D2}), leading to the selection of the XY sequence for the experiment, as depicted in Figure \ref{FigNVNV}(c).
The simulations also suggest a practical approach to ensure that \(N\) is sufficiently large to satisfy the conditions of Average Hamiltonian Theory (AHT) without prior knowledge. 
This involves performing several measurements with different values of \(N\). 
When the measured decay time no longer changes with \(N\) (i.e., saturates), the saturated \(T_2\) can be attributed to the effective Hamiltonian containing only the NV-NV interaction term.
For the example sample, as shown in Figure \ref{FigNVNV}(c-e), the decoherence rate saturates at $\Gamma_2 = 3\ \unit{kHz}$ when the repetition time in the XY8 sequence is sufficiently large. Consequently, $\Gamma_\text{NV-NV}$ is determined to be $3\ \unit{kHz}$, corresponding to an NV concentration of approximately $0.17\ \unit{ppm}$.

Interestingly, we notice that the decay rate increases with \textit{N} in the simulation but the decay rate decreases in the experiment.
In the simulations, higher-order terms in the Magnus expansion of the effective Hamiltonian obscure part of the NV-NV interaction strength.
As the repetition number \textit{N} increases, the effective Hamiltonian approaches the first-order term, recovering the full strength of the NV-NV interaction.
Consequently, the decay rates measured using the XY8-\textit{N} sequence increase with \textit{N} in the simulations.
However, in the experiments, neglected AC noise has a greater impact on the decay rate than the discrepancies caused by higher-order terms observed in the simulations.
Sequences with larger repetition numbers \textit{N} exhibit narrower bandwidths, effectively filtering out AC noise. 
As a result, the measured decoherence time increases with \textit{N}, as shown in Figure \ref{FigNVNV}(d) and (e).
In either case, $\Gamma_{\text{NV-NV}}$ can be reliably estimated using a large repetition number \textit{N} in the XY8-\textit{N} or XY16-\textit{N} sequence.

It is important to note that due to the limit imposed by $\Gamma_1$, the detectable NV-NV interaction strength must exceed $0.1\ \unit{kHz}$, corresponding to an NV concentration of approximately $10\ \unit{ppb}$.
Regarding such samples, comparing the fluorescence intensity with a known sample could be one of the most effective ways to determine concentrations.

\section{\label{Sec III}Results and Discussions}
	
\begin{table}[b]
	\centering
	\caption{The measured decay rates of the samples}
	\label{tab_measurementsresult}
	\begin{ruledtabular}
	\begin{tabular}{cccccccc}
		\multirow{2}{*}{No.}& {$\Gamma_2^*$} & {$\Gamma_2$} & {$\Gamma_\text{DQ/2}$} & {$\Gamma_\text{elec}$} & {$\Gamma_\text{strain}$} & {$\Gamma_\text{C13}$} & {$\Gamma_\text{NV-NV}$} \\
		& {(kHz)} & {(kHz)} & {(kHz)} & {field (kHz)} & {(kHz)} & {(kHz)} & {(kHz)} \\
		\hline
		\multirow{2}{*}{H1}&{165(1)}&\multirow{2}{*}{4.5}&\multirow{2}{*}{125}&\multirow{2}{*}{2.9}&\multirow{2}{*}{34.6}&\multirow{2}{*}{$\sim$30}&\multirow{2}{*}{2}\\
		&{137(1.5)}\\
        \addlinespace
		\multirow{2}{*}{H2}&{129(1)} &\multirow{2}{*}{10}&\multirow{2}{*}{108}&\multirow{2}{*}{4.4}&\multirow{2}{*}{44.6}&\multirow{2}{*}{$\sim$30}&\multirow{2}{*}{3}\\
		&{117(1.2)}\\
        \addlinespace
		H3 & 581     & 30.0  & 595 & 6.5 & 34.5  & $\sim$5  & 5 \\
        \addlinespace
		H4 & 385     & 32.0  & 156 & 7.2 & 160.8 & $\sim$5  & 6.6 \\
        \addlinespace
		H5 & 741     & 109.0 & 405 &11.8 & 253.2 & $\sim$5  & 13.3 \\
        \addlinespace
		H6 & 2800    & 625.0 & 1720& 25.5& 274.5 & $\sim$5  & 62 \\
        \addlinespace
        \hline
        \addlinespace
		\multirow{2}{*}{C1} &{96(1)}  &\multirow{2}{*}{4.2}&\multirow{2}{*}{86}&\multirow{2}{*}{2.3}&\multirow{2}{*}{30.6}&\multirow{2}{*}{$\sim$1}&\multirow{2}{*}{1.7}\\
		&{85(1.3)}\\
        \addlinespace
		C2 & 167    & 8.0    & 77.5& 3.5 & 88.5  & $\sim$1  & 2.1 \\
        \addlinespace
		\multirow{2}{*}{C3}&\multirow{2}{*}{800}&{15(1)} &\multirow{2}{*}{96}&\multirow{2}{*}{6.2}&\multirow{2}{*}{1273.8}&\multirow{2}{*}{$\sim$1} &\multirow{2}{*}{3}\\
		&&{27(0.5)}\\
        \addlinespace
		C4 & 714    & 70.0   & 481 & 19  & 272   & $\sim$1  & 20 \\
        \addlinespace
		C5 & 555    & 83.0   & 277 & --- & 363.0 & $\sim$1  & 22 \\
	\end{tabular}
    \end{ruledtabular}
\end{table}

\begin{figure*}[b]
    \includegraphics{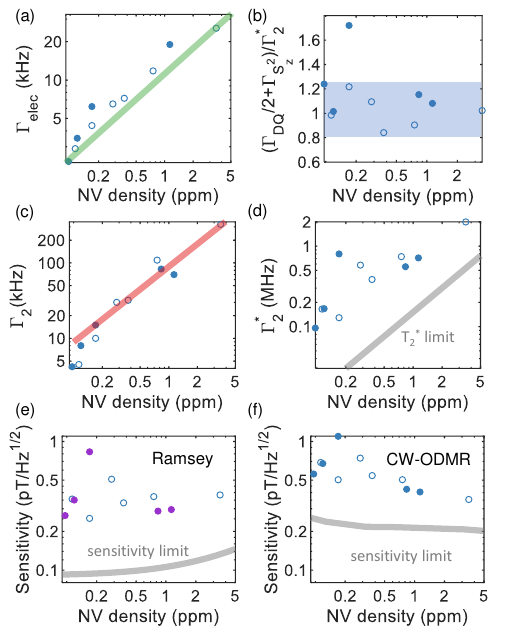}
    \caption{Characterization results from 11 diamond samples, where solid and hollow markers represent CVD- and HPHT-grown diamonds, respectively. (a) Decay rates induced by intrinsic electric field noise as a function of NV concentration, with the line showing the theoretical lower bound. (b) Ratio between $\Gamma_2^*$ and $\Gamma_{\text{DQ}}/2+\Gamma_{S_z^2}$ (sum of dipolar-induced and $S_z^2$-related decay rates), where the shaded area represents deviations within 20\%. (c) $\Gamma_2$ measured using Hahn-Echo sequence, with the red line showing the fitted linear relationship. (d) $\Gamma_2^*$ measurements with an estimated lower bound achievable through strain noise elimination. Calculated sensitivities based on measured $\Gamma_2^*$ values and theoretical limits using (e) Ramsey method and (f) CWODMR.}
    \label{Fig_12sample}
\end{figure*}

Using the presented methods, we analyzed the dephasing characteristics of two groups of diamonds: six samples grown using the high-pressure-high-temperature (HPHT) method (labeled H1–H6) and five samples grown using the chemical vapor deposition (CVD) method (labeled C1–C5).
The initial nitrogen concentrations of the samples ranged from approximately 1 ppm to 30 ppm. The samples were irradiated and annealed under various conditions to optimize the conversion rate of NV centers (details in Appendix \ref{Appendix E}).
The results of the dephasing measurements are summarized in Table \ref{tab_measurementsresult}. Notably, the example sample described in Section \ref{sec II} corresponds to H2 in Table \ref{tab_measurementsresult}.
The longest $T_2^* = 11.8 \unit{\mu s}$ was measured in Sample C1, which has the lowest estimated NV center concentration and strain among all the samples.
Most samples exhibited a single exponential decay in the $T_2^*$ measurements, except for a few with relatively long $T_2^*$ values, namely H1, H2, and C1.
A stretched exponential decay curve was used to fit the $T_2^*$ of the three samples. The stretch factors are noted in parentheses in the table, and the dephasing rates obtained by forcing a single exponential decay fit are also reported.

To visualize the results, we plot different decay rates as functions of NV center concentration, as shown in Figure \ref{Fig_12sample}.
The NV center concentrations of the samples were determined by measuring NV-NV interactions, as described in Section \ref{sec IID}.
Since the strain distribution in diamonds strongly depends on the crystal growth process, $\Gamma_{\text{strain}}$ is expected to be independent of NV center concentration. 
Therefore, $\Gamma_{\text{elec}}$ is plotted instead of $\Gamma_{\text{strain}}$ in Figure \ref{Fig_12sample}(a), even though the $S_z^2$ term in ${\cal{H}}_{NV}$ is primarily induced by inhomogeneous strain rather than the electric field.
As shown in the figure, $\Gamma_{\text{elec}}$ exhibits a strong correlation with NV center concentration. These results confirm that the electric field noise in diamonds primarily originates from NV centers and their associated electron donors, as described in reference \cite{46PhysRevLett2018Yao}. The predicted scaling factor of $\Gamma_{\text{elec}}$ with [NV] is 560 $\unit{kHz/ppm^{2/3}}$, as indicated by the green line in the figure.
Furthermore, the results suggest that the upper limit of NV center concentration can be estimated based on measurements of electric field noise in diamonds.

As described in Section \ref{sec IIA}, the dephasing rate $\Gamma_2^*$ can be approximated as $\Gamma_\text{DQ/2} + \Gamma_{S_z^2}$. To evaluate this approximation, Figure \ref{Fig_12sample}(b) plots the ratio $(\Gamma_\text{DQ/2} + \Gamma_{S_z^2}) / \Gamma_2^*$. Most ratios fall within a 20\% error margin, except for Sample C3, where $\Gamma_2$ is better fitted with a stretch factor of $p = 0.5$, likely due to the high strain-induced decay rate in this sample. These results suggest that the independent-noise-source summation model generally describes dephasing behavior but may oversimplify interactions between different noise sources, such as magnetic field noise competing with strain or electric field noise, as discussed in Ref. \cite{RN48}.

The Hahn-echo measurements of $\Gamma_2$, shown with a linear fit in Figure \ref{Fig_12sample}(c), reveal a strong correlation with NV center concentration.
$\Gamma_2$ is mainly contributed by: 1. the NV-NV interactions, and 2. the AC components of the P1 center spin bath.
Both sources contribute linearly to the decay rate, depending on their concentrations.
With similar conversion efficiencies from the P1 centers to the NV centers for all samples, density of NV center, initial N, and residual P1 center should be correlated (sample information details in Appendix \ref{Appendix E}), which leads to correlations observed in Figure \ref{Fig_12sample}(c).
Small deviations from the fit for a few samples can be attributed to variations in the conversion efficiencies from P1 centers to NV centers (details in Appendix \ref{Appendix E}).

Since the sensitivity per volume of DC diamond magnetometry depends directly on $\Gamma_2^*$ and NV center concentration, the data are plotted in Figure \ref{Fig_12sample}(d). 
In general, the measured $\Gamma_2^*$ increases with NV center concentration, though the correlation is not strictly linear. 
This behavior can be explained using the mean-field model, which predicts that spin-related decay rates scale linearly with concentration, while discrepancies arise from contributions of $\Gamma_\text{strain}$.
Strain noise could potentially be reduced with advancements in crystal growth technology. Using the measured data, we estimate the achievable limits of $\Gamma_2^*$ under improved growth conditions, assuming strain can be further eliminated and dipolar terms establish a linear relationship between density and $\Gamma_2^*$, as indicated by the gray line in Figure \ref{Fig_12sample}(d).

Finally, we evaluate the magnetic field sensitivity achievable by the samples based on their dephasing properties. Ramsey and continuous-wave ODMR are the two commonly used methods for DC magnetic field sensing. The maximum fluorescence photon number is assumed to scale linearly with NV density. This assumption, along with $T_2^*$ and the estimated NV concentrations, is used to calculate the shot noise limits based on the Ramsey method \cite{22RevModPhys2020Walsworth}. For CW-ODMR, a five-level rate equation model is applied to determine the sensitivity limit \cite{RN33}.
In the simulations, a volume of 0.04 $\unit{mm^3}$ was set for all diamonds. The results are shown in Figure \ref{Fig_12sample}(e) and (f). Additionally, the optimal sensitivity limits, derived from the optimal $\Gamma_2^*$ relation to NV density(gray line in Figure \ref{Fig_12sample}(d)), are plotted for both Ramsey and CWODMR.

Understanding the dephasing characteristics of NV center ensembles provides a pathway to improve sensitivity by extending $T_2^*$. Improving dephasing times relies heavily on both the growth process and defect treatment of diamonds. Low-strain diamonds, typically produced using the HPHT technique, are particularly critical. Recently, advancements in the CVD method have enabled the growth of low-strain diamonds by using low-strain HPHT diamonds as substrates, as demonstrated by samples C1 and C2 in Table \ref{tab_measurementsresult}.
Additionally, techniques to suppress specific dephasing sources, as indicated by our results, are necessary.
For instance, the double-quantum manipulation technique combined with bath driving effectively suppresses both strain and P1 center bath noise \cite{9PhysRevX20185Walsworth}.
Residual dephasing from NV-NV interactions can also be mitigated using dipole-dipole decoupling sequences, such as WAHUAHA and MREV8, which average out the NV-NV interaction term in the effective Hamiltonian \cite{RN25}.
Experimentally, the DC sensitivity determined by the $\Gamma_2^*$ is a few $\unit{pT/\sqrt{Hz}}$ for the 0.04$\unit{ mm^3}$ diamond. This is still an order of magnitude away from the shot-noise limit.
This discrepancy is primarily attributed to technical noise, such as laser noise, which is challenging to reduce further at the current stage.
Thus, extending $T_2^*$ through advanced techniques represents the most effective approach to enhancing DC sensitivity.
Unraveling the dephasing mechanisms of each sample enables the combination of multiple noise-decoupling techniques to further extend $T_2^*$.

\section{Conclusions and outlooks}
This work proposes and demonstrates multiple techniques to characterize and understand the dephasing mechanisms of NV center ensembles in diamonds.
A diverse set of diamonds, grown and treated under varying conditions, were experimentally studied.
Dephasing sources were classified into three groups, and their corresponding decay rates were measured.
Intrinsic strain and electric-field noise were precisely characterized using a combination of the strain-CPMG sequence and zero-field ODMR spectrum.
The decay rate from dipolar noise, primarily due to the P1 center spin bath, was measured using the duration-sweep DEER method for samples with low paramagnetic defect concentrations.
For NV-NV interactions, the XY8-N sequence was proposed to resolve weak NV-NV couplings and NV concentrations.
More importantly, understanding the dephasing sources for a given sample provides guidance for selecting specific techniques to suppress dominant noise sources, thereby enhancing DC magnetic field sensitivity. Discussions are targeted to the sensitivity limits that could be achieved with the samples used in this work.

In addition to the techniques discussed, further improvements are expected with advancements in diamond growth technology. For instance, phosphorus-doped n-type diamonds have demonstrated dephasing times exceeding 1 ms for single NV centers \cite{RN68}.
One or two orders of enhancement on the $\Gamma_2^*$ for NV center ensemble in the n-type diamond can be expected.
By improving the DC magnetic field sensitivity to the sub-pT level, diamond magnetometry can be competitive with the state-of-the-art highly-sensitive quantum magnetometers while maintaining sub-mm spatial resolution.

\section{Acknowledgement}
Jixing Zhang and Cheuk Kit Cheung contributed equally to this work as co-first authors. We acknowledge the financial support by EU via project AMADEUS, the BMBF via projects DiaQnos and NeuroQ as well as QHMI and the DFG via GRK 2642. In addition, C.Z. acknowledges the financial support by the NSFC via Grant 62473058. The work was performed in part under the collaboration agreement among University of Tsukuba, University of Ulm and University of Stuttgart.

\appendix

    \section{\label{Appendix A}Details of strain and electric field Hamiltonian}
    The Hamiltonian describing the effects of strain and electric fields on the electron spin of an NV center is given by:
    \begin{multline}
        {\cal{H}}_{\text{strain+ef}}=(\mathcal{M}_z+d^\parallel E_z) S_z^2+(\mathcal{M}_x+d^\perp E_x)(S_y^2-S_x^2) +(\mathcal{M}_y+d^\perp E_y)(S_xS_y+S_yS_x)
    \end{multline}
    Here, $d^\parallel = 0.35\ \unit{Hz\cdot cm/V}$ and $d^\perp = 17\ \unit{Hz\cdot cm/V}$ denote the parallel and transverse electric dipole moments~\cite{RN39}, while $E_{x,y,z}$ represent the components of the electric field.
    Terms like $S_xS_z$ are neglected due to the dominance of the zero-field splitting term $D_{gs}$.
    The spin-strain coupling amplitudes $\mathcal{M}{x,y,z}$ are related to the strain tensor $[\epsilon{ij}]_{i,j=x,y,z}$ as follows:
    \begin{equation}
	\begin{aligned}
		& \mathcal{M}_x=b\left(\epsilon_{x x}-\epsilon_{y y}\right)+2 c \epsilon_{y z}, \\
		& \mathcal{M}_y=-2 b \epsilon_{x y}-2 c \epsilon_{x z}, \\
		& \mathcal{M}_z=a_1 \epsilon_{z z}+a_2\left(\epsilon_{x x}+\epsilon_{y y}\right),
	\end{aligned}
    \end{equation}
    Here, $a_1 = -8\ \unit{GHz/strain}$, $a_2 = -12.4\ \unit{GHz/strain}$, $b = -3.7\ \unit{GHz/strain}$, and $c = 11.8\ \unit{GHz/strain}$ are the spin-strain coupling parameters \cite{RN45, RN47, RN49,51PhysRevB2019walsworth}.
    For a single NV center, strain and electric fields induce shifts in energy levels. When a magnetic field $\Vec{B}0$ is applied, the effects of $E{x,y}$ and $\mathcal{M}_{x,y}$ are reduced by a factor $\sqrt{(d^{\perp} E_x + \mathcal{M}_x)^2 + (d^{\perp} E_y + \mathcal{M}_y)^2}/(\gamma_e B_z)$.
    Therefore, under a 10 Gauss bias magnetic field, only $\mathcal{M}_z+d^\parallel E_z$ contributes to the energy level shifts. 
	
    For the NV center ensemble case, the random distribution of electric field and strain induce a distribution of energy levels, which leads to dephasing.
    To simplify the analysis, the distributions of strain and electric fields are assumed to be isotropic, with each component following a Gaussian distribution centered at zero and standard deviations $\sigma_\epsilon$ and $\sigma_E$, respectively.
    Under these assumptions, the dephasing contributions are given by $\Gamma_{\text{strain}}=(a_1+2a_2)\sigma_\epsilon$ and $\Gamma_{\text{elec}}=d^\parallel \sigma_E$. Consequently, the decay measured by interferometry sequences is:
    \begin{equation}\label{eq_strain2}
        \Gamma_{S_z^2}= {\Gamma_{\text{strain}}}+{\Gamma_{\text{elec}}} =(a_1+2a_2)\sigma_\epsilon +d^\parallel \sigma_E
    \end{equation}
    Similarly, the zero-field ODMR dip linewidth is given by:\cite{46PhysRevLett2018Yao}
	
    \begin{equation}\label{eq_strain3}
	\nu_{\text{zf-ODMR}} = d^{\perp} \sigma_E.
    \end{equation}
    Using Eq.~\ref{eq_strain2} and Eq.~\ref{eq_strain3}, the values of $\sigma_\epsilon$ and $\sigma_E$ and their respective contributions to dephasing are determined.	
    Electric field noise arises from uniformly distributed defects carrying both positive and negative charges within the diamond.
    A model has been established to quantify the relationship between charge defect density and zero-field ODMR dip linewidth, given as $0.56\ \unit{MHz/ppm^{2/3}}$ \cite{46PhysRevLett2018Yao}.    
    Given that the NV$^-$ center carries a negative charge, the NV$^-$ concentration in the example sample discussed in Section \ref{sec IIA} is estimated to be below 0.2 ppm.
    The transverse component along a specific axis can be isolated by applying a transverse magnetic field, selectively retaining it in the desired direction \cite{50nanolett.9b00900}.   
    Electric and stress field information can also be obtained with high sensitivity through Photoluminescence Excitation (PLE) measurements in the excited state.
    However, this approach requires cryogenic measurements, which is outside the scope of this study.
	
    \section{\label{Appendix B}Model of ensemble spin-spin interaction based on mean-field theory}

    The dipolar interaction Hamiltonian between a single bath spin (denoted as B-spin) and an NV center is given by:
    \begin{equation}
        {\cal{H}}_\text{inter} = \frac{\mu_0\gamma_{B}\gamma_{NV}\hbar^2}{4\pi|\vec{r}|^3}(-3(\vec{S}_{B}\cdot\hat{r})(\vec{S}_{NV}\cdot\hat{r})+(\vec{S}_{B}\cdot\vec{S}_{NV})).
	\label{eq_dipoledipole}
    \end{equation}
    In this equation, $\vec{S}B$ and $\vec{S}{NV}$ represent the spins of the B-spin and NV center, respectively. 
    $\vec{r}$ is the distance vector between the two spins, and $\gamma_B$ and $\gamma_{NV}$ are their gyromagnetic ratios.
    Given the substantial zero-field splitting of the NV center, the off-diagonal terms in the interaction Hamiltonian can be neglected (secular approximation), leaving only the terms $\left\{S_{NV}^zS_{B}^z, S_{NV}^zS_{B}^x, S_{NV}^zS_{B}^y\right\}$.
    Furthermore, since the energy splitting of the bath spin is also much larger than the interaction strength, only the term $\left\{S_{NV}^zS_{B}^z\right\}$ is left.

    \begin{figure}[b]
	\centering
	\includegraphics{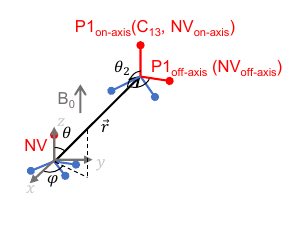}
	\caption{Schematic representation of the interaction between an NV center and a single spin from the spin bath}
	\label{FigP1-NVposition}
    \end{figure}	
 
    We assume the B spin is a spin-$\frac{1}{2}$ electron spin coupled to a spin-$M$ nuclear spin, and the NV center is surrounded by $N$ non-polarized B-spins.
    One can define the interaction strength from each hyperfine level of B spin as $A_k^i=S_{NV}^zS_{B}^z$, which can be calculated as follows,
   \begin{equation}
	A_k^i=\lambda^i_{H_{NV}+H_{B,k}+H_\text{inter,k}}-\lambda^i_{H_{NV}+H_{B,k}}.
	\label{eq_P1energyDiff}
    \end{equation}
    Here, $\lambda^i$ is the $i$th eigenvalue of the corresponding Hamiltonian.
    The free induced decay (FID) of a single NV center is expressed as~\cite{28PhysRevB2008Hanson}, 
    \begin{equation}
	M_x^i(t)=\frac{1}{4M+2}\sum^{4M+2}_{i}\prod_{k}^{N}\cos\frac{A^i_{k}t}{2}.
	\label{eq_singledephasing}
    \end{equation}
    
    When $N$ is sufficiently large, the central limit theorem can be applied, and the FID curves distribution approximately follows the Gaussian,
    \begin{equation}
	M_x(t)\overset{N\rightarrow\infty}{=}1/2(1+e^{-(t/T_2^*)^2}).
    \end{equation}
    Here, $T_2^* = \sqrt{2 / \sum_i{\Delta_i^2}}$ represents the dephasing time, where $\Delta_i^2 = \sum_k{{A_k^i}^2}.$
	
    For NV center ensembles, it is assumed that the term $\sum_i{\Delta_i^2}$ for each NV center is independent and that both NV centers and B spins are homogeneously distributed throughout the diamond.
    The probability of a certain distribution of $N$ B-spins is $\prod_{k}^{N}dV_k/V$, where $V$ is the volume of the sample and $dV_k$ is the volume element that the $k$th B-spin is located at. 
    At room temperature, the B-spins are fully thermalized.
    First, We calculate the contribution of all B-spins at the $i$th energy level to the dephasing of the NV ensemble.
    The overall contribution can be obtained by the sum of $M_x^i$, following Equation \ref{eq_singledephasing}.
    The FID curve for the ensemble NV center is obtained by integrating over all possible single NV center FID signals. By setting $U^i_k(\theta, \phi) = r^3 A^i_k$, we derive:
    \begin{equation}\label{}
	\begin{split}
		M^i_x(t) &= V^{-N}\int_V dV_1 \cdots \int_V V_N \cos{A^{k,i}t}\\
		&=(V^{-1}\int_V \cos{A^i_{k}t}dV)^N\\
            &=(V^{-1}\int_{-1}^1 d\cos{\theta}\int_{-\pi}^{\pi} d\phi \int_0^\infty r^2 \cos{(U^i_{k}t r^{-3})}dr)^N.
	\end{split}
    \end{equation}
    Substituting $r$ with $x$ yields:
    \begin{equation}\label{}
	\begin{split}
		\int_0^\infty r^2 \cos{(U^i_{k}t r^{-3})}dr=
		\frac{1}{3}\int_{-\infty}^0 x^{-2} \cos{(U^i_{k}t x)}dx\\
            = \frac{V}{4\pi}-\int_{-\infty}^0 \frac{1-\cos{(U^i_{k}t x)}}{x^{-2}}dx=\frac{V}{4\pi}-\frac{\pi|U^i_{k}t|}{2}.
	\end{split}
    \end{equation}
    When $N \rightarrow \infty$ and $V \rightarrow \infty$, the concentration of B-spin, $n_B=N/V$, remains finite and the FID signal is derived as~\cite{30ConceptsofMagneticResonance1997Romanelli},
    \begin{equation}
	\begin{split}
            M^i_x(t)&=\lim\limits_{N,V \to \infty}(1-V^{-1}\int_{-1}^1 d\cos{\theta}\int_{-\pi}^{\pi} d\phi \frac{\pi|U^i_{k}t|}{2})^N\\ 
		&=\exp(-\int_{-1}^1 d\cos{\theta}\int_{-\pi}^{\pi} d\phi \frac{\pi|U^i_{k}t|}{2}\cdot n_{B_i}).
	\end{split}
    \end{equation}
    Taking all energy levels into account, the dephasing rate induced by a B spin is given by:
    \begin{equation}\label{eq_BspinDephasing}
	\Gamma_{B}=\sum_i{\int_{-1}^1 d\cos{\theta}\int_{-\pi}^{\pi} d\phi \frac{\pi|U^i_{k}t|}{2}\cdot n_{B_i}}.
    \end{equation}
    When the quantum principal axes of the B spin and the NV center align, and the B spin is spin-$\frac{1}{2}$, the integrals over $\theta$ and $\phi$ yield an analytical solution:
    \begin{equation}\label{eq_FID}
	M_x(t,\phi)=\exp[{-(4\pi^2 /9\sqrt{3})\mu_0 \frac{\gamma_{B}\gamma_{NV}\hbar}{4\pi}n_B t}].
    \end{equation}
    Here, $\Gamma_B/n_B=(4\pi^2 /9\sqrt{3})\mu_0 \frac{\gamma_{B}\gamma_{NV}\hbar}{4\pi}=141$ kHz/ppm.
    With 1 ppm corresponding to a density of $1.76 \times 10^{17} \text{cm}^{-3}$ in diamond, the P1 center concentration can be calculated accordingly when the bias field is sufficiently large.
    When the bias field is weak, Eq. \ref{eq_BspinDephasing} is employed to obtain a numerical solution for $M_x$, as no straightforward analytical solution is available.
	
    When the pulse sequences include MW $\pi$ pulses to drive the NV centers and RF $\pi$ pulses to drive the P1 centers, $|U^i_{k}t|$ is replaced by:  
    \begin{equation}\label{eq_FinalMeanFiel}
        \sum_{i \neq n, m} \left|\int_t s_{NV}(t) U_k^i t \, dt\right| + \left|\int_t s_{NV}(t) \left(r_n(t) U_k^n t + r_m(t) U_k^m t\right) dt\right|.
    \end{equation}
    Here, $s_{NV}(t)$ represents the gate function with values of $-1$ or $1$, determined by the MW $\pi$ pulses in the NV sequence, while $r_n(t)$ denotes the gate function determined by the RF $\pi$ pulses in the bath spin sequence.
    This framework serves as a generalized mean-field approach for calculating decay rates induced by spin ensembles.
    The model is subsequently applied to analyze the decay rates induced by P1 centers.

    Moreover, the operation of taking the absolute value in Eq. \ref{eq_BspinDephasing} ensures that the ensemble decay is independent of the polarization of B spins, indicating that this method is applicable to both off-axis NV centers (polarized by laser) and P1 centers (unpolarized).
	
    \section{\label{Appendix C}Details of measuring P1 center dephasing contribution under week field}
	
    \subsection{\label{Appendix C1}P1 Hamiltonian and model of the decay in DEER measurements}
    The Hamiltonian for a P1 center is expressed as:
    \begin{equation}
        H_{\text{P1}} = \gamma_{e}\vec{S}^{\text{p1}}\vec{B_0} + A_{\parallel}S_{z}^{\text{p1}}I_{z}^{\text{p1}} + A_{\perp}(S_{x}^{\text{p1}}I_{y}^{\text{p1}}+S_{y}^{\text{p1}}I_{x}^{\text{p1}}) + Q(I^{\text{p1}}_z)^2.
	\label{eq_P1hamiltonian}
    \end{equation}
     Here, $A_\parallel=114.2 \text{MHz}$ and $A_\perp=81.8 \text{MHz}$ are the hyperfine couplings strengths, $Q=3.97 \text{MHz}$ is the nuclear quadrupole, $S^{\text{p1}}$ and $I^{\text{p1}}$ denote the electron spin and nuclear spin of P1 center. 
    The Zeeman splitting of the nuclear spin is neglected because its magnitude is negligible compared to other terms in the Hamiltonian.
    The P1 center corresponds to the neutral charge state of single substitutional nitrogen ($N_s^0$), with an unpaired electron spin $S = 1/2$.
    The unpaired electron occupies the anti-bonding orbital of one of four N-C bonds, with about 25\% of the spin density localized on the nitrogen and about 75\% on the unpaired carbon dangling bonds. The antibonding character is lowered by a structure relaxation of elongating the N-C bond by about 30\% from the normal C-C bond. With the $C_3V$ symmetry, the P1 center has four orientations in the host diamond lattice, which under the B0 field along [111] can be separated into on-axis and off-axis categories. Due to the transverse coupling of the nuclear spins, the eigenstates of P1 centers are a mixed state of the electron spin, permitting several forbidden transitions.
	
\begin{figure}[t]
	\includegraphics{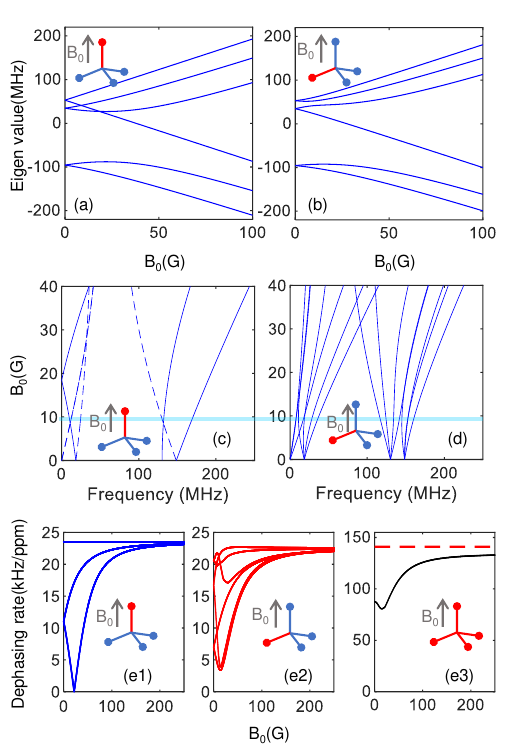}
        \caption{ The energy levels for on-axis(a) and off-axis(b) P1 centers as a function of magnetic fields. The inset shows the relative orientation between the P1 axis (red bar) and magnetic field (gray arrow). The transition frequency (DEER resonance peaks) for the on-axis(c) and off-axis(d) P1 centers versus magnetic fields. The light blue line corresponds to the 9.5 G magnetic field which is the same as the experimental condition in Figure \ref{figP1}(a).
        Dephasing rates per unit concentration from different eigenstates of the on-axis(e1) and off-axis(e2) P1 center and the total dephasing rates per unit concentration from P1(e3).}
	\label{FigNVP1NEW}
    \end{figure}

    Using the Hamiltonian of the P1 center, we calculate its eigenstates and eigenvalues under various magnetic fields, as illustrated in Figure \ref{FigNVP1NEW}(a) and (b).
    In the case of an on-axis P1 center under a weak magnetic field, strong transverse coupling between electron and nuclear spins gives rise to several forbidden transitions, as depicted in Figure \ref{FigNVP1NEW}(c).
    For the off-axis P1 center, the misalignment of the magnetic field introduces more forbidden transitions.
    The following method can be used to verify whether the transition is allowed.
    The microwave-driving Hamiltonian is expressed as:
    \begin{equation}
	H_{P1_{RF}}(t)=(B_xS_x^{P1}+B_yS_y^{P1}+B_zS_z^{P1})\cos(\omega t).
    \end{equation}
    The eigenstates of the P1 center are denoted as $\vert \phi_m \rangle$ ($m = 1, \ldots, 6$). A transition occurs when $\langle \phi_n \vert H_{P1_{RF}}(t) \vert \phi_m \rangle \neq 0$.
    Using this method, the DEER spectrum under varying magnetic fields is simulated, as illustrated in Figure \ref{FigNVP1NEW}(d).

    Furthermore, the relationship between the P1 center density and the DEER decay corresponding to a specific P1 transition is calculated using the mean-field theory presented in Appendix B.
    For the sweep pulse DEER sequence, the gate function used in eq. \ref{eq_FinalMeanFiel} is:
    \begin{equation}\label{}
	\begin{split}
		s_{NV} &= 1, \quad t \in (0, T_\text{fix})\\
		&= -1, \quad t \in (T_\text{fix},2T_\text{fix})\\
		s_{n}(t)(s_{m}(t)) &= 1(0), \quad t\in (0,2T_\text{fix}-T)\\
		&= 0(1), \quad t \in (2T_\text{fix}-T,2T_\text{fix}).
	\end{split}
    \end{equation}
    Then, the decay rate can be calculated using eq.\ref{eq_BspinDephasing} and eq.\ref{eq_FinalMeanFiel} as, $\Gamma_\text{P1,DEER}/n_\text{P1}= 11 \text{kHz/ppm}$ at the resonance peak at 152 MHz @ 9.5 G that we employed.
    The pulses of NV and P1 are not perfect in the actual experiment. As demonstrated in ~\cite{1PhysRevMaterials2019Jayich,3PhysRevB2021Budker}, the spin can only be fully flipped when the Rabi frequency approaches $+\infty$.
    Thus, based on the method described in~\cite{1PhysRevMaterials2019Jayich}, the relationship is revised to $\Gamma_\text{P1, DEER}/n_\text{P1}= 13 \, \text{kHz/ppm}$.
    Consequently, the P1 center concentration in the example sample is determined to be 0.2 ppm.

    The relation between P1 density and $\Gamma_P1$ is also calculated using eq.\ref{eq_BspinDephasing}.
    Figure \ref{FigNVP1NEW}(e) illustrates the contributions of the six energy levels of P1 centers to the dephasing of NV centers. These contributions are shown separately for the on-axis (Figure \ref{FigNVP1NEW}(e1)) and off-axis (Figure \ref{FigNVP1NEW}(e2)) P1 groups, assuming unit concentration.
    The total contributions of the P1 centers to the dephasing of NV under different magnetic fields are obtained by summing the six energy levels of the two groups according to their proportions (equal distribution at room temperature).
    In Figure \ref{FigNVP1NEW}(e3), the red dashed line corresponds to the conventional calculation method of $K=$141 kHz/ppm, which represents the upper limit of dephasing generated by a general electron bath spins on the NV center through $S_zS_z$  term.

    \subsection{\label{Appendix C2}P1 bath driving with DEER pulse}
    \begin{figure}[t]
        \includegraphics{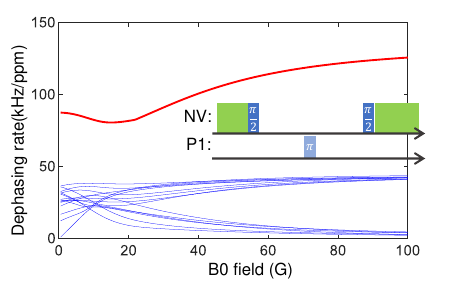}
        \caption{Residual dephasing rates for P1 center after 15 possible spin bath driving combinations(blue) compared with the bare P1 dephasing(red) as a function of magnetic field.}
	\label{Figbathdriving}
    \end{figure}
    Due to the spin-1 feature of the host nitrogen nuclear spin in the P1 center, 6 sets of microwave frequencies are required to drive all hyperfine transitions for the spin bath driving so that both on- and off-axis P1 centers can be addressed.
    In the bath driving method for P1 centers, interactions from on-axis P1 centers, which exhibit high symmetry under weak magnetic fields, can be completely suppressed using existing spin bath driving methods.
    However, this method is not effective in driving the off-axis P1 centers. 
    To address this limitation, all possible transition pairs for the three sets of microwave frequencies were analyzed, yielding $C(6,2) \times C(4,2) / (3!) = 15$ combinations.
    Figure \ref{Figbathdriving} shows the dephasing rates measured using the bath-driving sequence with all possible frequency combinations, plotted as a function of the $B_0$ field.
    At approximately 10 Gauss, the lowest residual dephasing rate after applying P1 bath driving is around 20 kHz/ppm, compared to 80 kHz/ppm without bath driving. This corresponds to a 4-fold reduction in dephasing.
    Similarly, we get an 11-fold improvement of the dephasing rate when B0 is 30 Gauss, and up to a 32-fold dephasing suppression can be achieved when the B0 is about 100 Gauss.

 \section{\label{Appendix D}Details of measuring NV-NV interactions}
	
    \subsection{\label{Appendix D1}NV DEER calculation details}

    Similar to the P1 center, the density of off-axis NV centers can be calculated using Eq. \ref{eq_BspinDephasing}.
    We take a similar method used in Ref. \cite{23PhysRevLett2018Lukin}.
    By setting the bias magnetic field in a certain direction, we split the resonant frequency of all four axes, and each corresponding microwave frequency address one of the hyperfine lines of each axis.
    The relationship between density and the NV DEER decay rate is given by $\Gamma_\text{NV, DEER}/n_\text{NV} = 10.1 \text{kHz/ppm}$. 

    \subsection{\label{Appendix D2}Simulation of dephasing 
    properties with different dynamics decoupling sequences}

    To identify the optimal sequence for accurately measuring the NV-NV interaction strength described in the main text, this section simulates the decay behavior of various sequences under different noise conditions. These results are compared against the ideal decay curve, which reflects only NV-NV interactions. The sequence that remains consistent with the ideal results, even in the presence of noise, is selected to ensure reliable and accurate measurements of NV-NV interaction strength.
    
    The simulation employs the Hamiltonian defined in Equation \ref{eq_dipoledipole}, which incorporates the interaction between two NV centers.
    A Monte Carlo method is applied to model randomly distributed disorders and interactions.
    The noise intensity is characterized by the standard deviation of these distributions.


    The simulation parameters are chosen to reflect realistic experimental conditions.
    First, the NV-NV interaction differs from a simple spin-1/2 system, adopting the form $S_x S_x + S_y S_y - S_z S_z$ \cite{23PhysRevLett2018Lukin}.
    A 10\% microwave pulse error, attributed to antenna inhomogeneity, is included in the simulations.
    Finite pulse durations are also incorporated into the simulations.
    The simulations use a disorder strength of $2\pi \times 1 \text{ MHz}$ and an interaction strength of $2\pi \times 10 \text{ kHz}$, derived from experimentally measured $T_2^*$ values and DEER interaction strengths.    
    This setup yields a disorder strength 100 times greater than the interaction strength, reflecting the maximum ratio observed in our measured samples. Thus, a sequence that satisfies these stringent conditions can be reliably applied to all samples in this study.
    It is important to note that AC field noise is excluded from the simulations to reduce computational complexity.

\begin{figure}[ht]
	\includegraphics{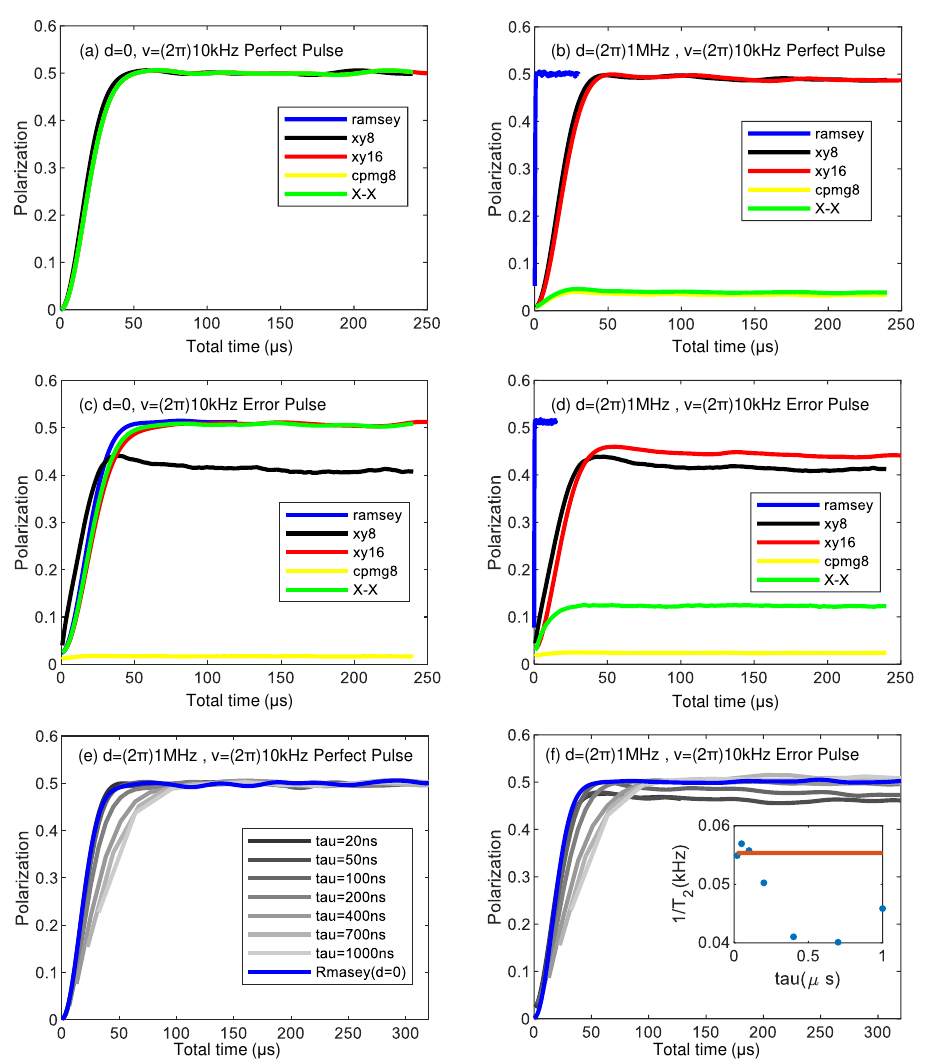}
        \caption{Simulation results of DD sequences with fixed \(\tau\) and increasing \(N\), where d denotes disorder strength and v represents NV-NV interaction strength. For (a-d), \(\tau = 50 \, \text{ns}\) with \(\pi\) pulse duration of 40 ns, under conditions specified in each panel. XY8 sequence simulations with varying \(\tau\) are shown without pulse error (e) and with pulse error (f). The blue curves represent Ramsey measurements under conditions matching panel (a). The convergence of XY8 results toward the blue curve indicates better measurement of the NV-NV interaction strength. Inset of (f) shows the fitted decoherence rates versus \(\tau\), where the red line indicates the decay rate from the Ramsey measurements under conditions matching panel (a)(NV-NV interaction limited decay rate).}
	\label{Fig_DDincreaseN}
	\end{figure}

\begin{figure}[ht]
	\includegraphics{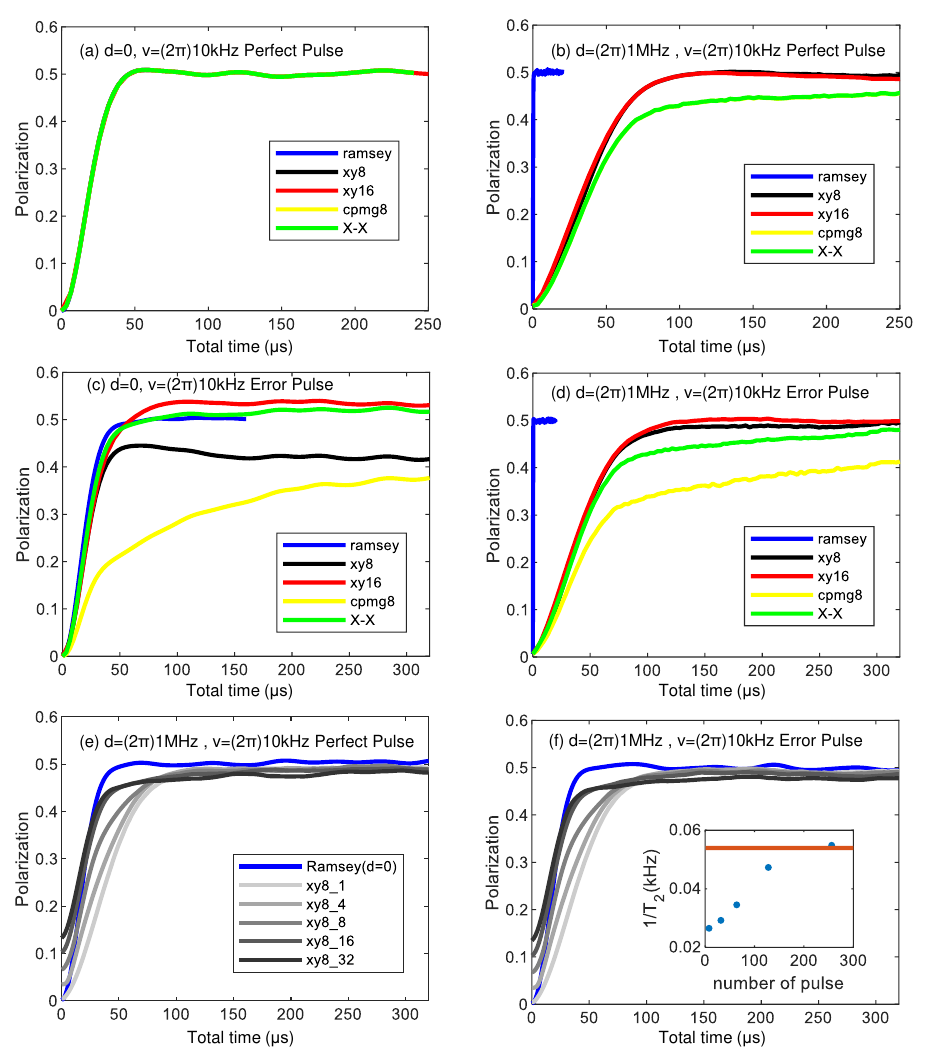}
        \caption{Simulation results of DD sequences with (N=1) and increase \(\tau\), where d denotes disorder strength and v represents NV-NV interaction strength. For panels (a-d), simulations use a \(\pi\) pulse duration of 40 ns under conditions specified in each panel. XY8 sequence simulations are shown without pulse error (e) and with pulse error (f). The blue curves represent Ramsey measurements under conditions matching panel (a). The convergence of XY8 results toward the blue curve indicates better measurement of the NV-NV interaction strength. Inset of (f) shows the fitted decoherence rates versus \(N\), where the red line indicates the NV-NV interaction limited decay rate.}
	\label{Fig_DDincreaseTau}
    \end{figure}

    The dynamic decoupling sequences simulated include XY8, XY16, CPMG, Ramsey, and the sequence described in Ref. \cite{RN64}, referred to here as X-X.
    As mentioned in the main text, there are two ways to measure \(T_2\).
    One way is to increase $N$, the number of $\pi$ pulses and fix pulse interval \(\tau\), and the other way is to fix $N$ and increase \(\tau\).
    Both methods are simulated, with results presented in Figures \ref{Fig_DDincreaseN} and \ref{Fig_DDincreaseTau}, respectively.

    We first investigated the sequence scheme where the pulse interval $\tau$ remains constant while increasing the pulse number \(N\).
    Under ideal conditions with only NV-NV interactions and no pulse errors, all sequences exhibit behavior similar to the Ramsey sequence, as demonstrated in Figure  \ref{Fig_DDincreaseN}(a).
    However, when introducing pulse errors (Figure \ref{Fig_DDincreaseN}(b)) and disorder (Figure \ref{Fig_DDincreaseN}(c)), the CPMG and X-X sequences show significant deviations from the original decay curve. Among all tested sequences, only XY8 and XY16 maintain fidelity to the ideal decay curve, leading to our adoption of the XY sequence scheme for subsequent experiments.
    Following the Average Hamiltonian Theory (AHT), which requires ${\cal{H}} \tau \ll 1$, we must carefully constrain $\tau$ to ensure valid AHT approximation. To optimize $\tau$, we performed simulations of the XY sequence scheme under both perfect(Figure 
    \ref{Fig_DDincreaseN}(e)) and error (Figure  \ref{Fig_DDincreaseN}(f)) pulse conditions. While both XY8 and XY16 sequences were simulated with nearly identical results, we present only the XY8 results for clarity. The inset in Figure \ref{Fig_DDincreaseN}(f) reveals that $\tau$ must be minimized to prevent significant deviations from true values.

    In the scheme where \(\tau\)is increased while the pulse number \(N\) remains constant, the behavior shown in Figure \ref{Fig_DDincreaseTau}(a) is consistent with the results from the \(N\)-increasing scheme.
    In the presence of pulse errors(\ref{Fig_DDincreaseTau}(b)) and disorder(\ref{Fig_DDincreaseTau}(c)), none of the sequences maintain consistency with the results in Figure \ref{Fig_DDincreaseTau}(a), and the CPMG sequence exhibits the most significant deviation.
    This is because, when increasing \(\tau\), ${\cal{H}}\tau$ gradually becomes larger until it no longer satisfies the AHT condition. 
    To satisfy this condition, we need to select a sufficiently large \(N\), ensuring that even as \(\tau\) increases, its value remains small enough to satisfy the AHT condition, given that the total time \(T = N \times \tau\) is fixed.
    Therefore, we tested different XY8-\(N\) sequences, in Figure \ref{Fig_DDincreaseTau}(e,f). 
    As depicted in the inset of Figure \ref{Fig_DDincreaseTau}(f), increasing \(N\) allows the XY8-\(N\) sequence to converge toward the true decay rate induced by the NV-NV interaction.

    Next, we examined the complementary sequence scheme where the pulse number \(N\) remains constant while increasing \(\tau\). The behavior observed in Figure \ref{Fig_DDincreaseTau}(a) initially shows consistency with the \(N\)-increasing scheme under ideal conditions. 
    However, introducing pulse errors (Figure \ref{Fig_DDincreaseTau}(b)) and disorder (Figure \ref{Fig_DDincreaseTau}(c)) causes all sequences to deviate from the ideal response, with the CPMG sequence showing the most pronounced deviation. 
    This deterioration occurs because increasing \(\tau\) causes ${\cal{H}}\tau$ to grow progressively larger, eventually violating the AHT condition. To maintain the AHT condition while allowing \(\tau\) to increase, we must choose a sufficiently large \(N\) such that \(\tau\) remains small enough within the fixed total time \(T = N \times \tau\). 
    We therefore investigated various XY8-\(N\) sequences, as shown in Figure \ref{Fig_DDincreaseTau}(e,f). 
    The inset of Figure \ref{Fig_DDincreaseTau}(f) demonstrates that increasing \(N\) enables the XY8-\(N\) sequence to converge to the true decay rate of the NV-NV interaction.
    
    According to the simulations, we find that both methods can be used to characterize the NV-NV interaction as long as \(\tau\) is sufficiently small or \(N\) is sufficiently large.
    In experiments, it is not straightforward to determine what value of \(\tau\) or \(N\) is sufficient for an unknown sample. 
    A simple solution is to experimentally replicate the simulations in Figure (f) by measuring multiple sets of curves, reducing \(\tau\) or increasing \(N\) until the measured decoherence time no longer changes.
    This approach is additionally beneficial by addressing the AC field noise, which we neglect in the simulations.
    Benefits come from the filtering effect in these DD sequences.
    Since the AC field noise follows a 1/f noise characteristic, the cancellation of the AC noises becomes more effective when $N$ is large.
    Simulation results are shown in the inset of Figure \ref{Fig_DDincreaseTau}(f) presenting the situation when AC noise is not considered. Here \(T_2\) decreases as \(N\) increases.
    On the other hand, the experimental results show that, for all the sample cases, \(T_2\) increases with $N$, indicating that the AC noise is gradually canceled.

    Our simulations demonstrate that both methods effectively characterize NV-NV interactions provided either \(\tau\) is sufficiently small or \(N\) is sufficiently large. However, determining optimal values of \(\tau\) or \(N\) for unknown samples presents an experimental challenge. We addressed this by systematically replicating the simulations shown in Figure \ref{Fig_DDincreaseTau}(f), measuring multiple sets of curves while increasing \(N\) until the measured decoherence time stabilized. This experimental approach offers an additional advantage in addressing AC field noise, which was not included in our simulations. The dynamic decoupling sequences provide an inherent filtering effect, particularly effective against the characteristic 1/f AC field noise when using large values of $N$. While our simulation results (inset of Figure \ref{Fig_DDincreaseTau}(f)) show \(T_2\) decreasing with increasing \(N\) in the absence of AC noise, our experimental measurements across all samples reveal an increase in \(T_2\) with increasing $N$, demonstrating progressive cancellation of AC noise effects.
    
    \section{\label{Appendix E}Additional information on the samples}
    
\begin{figure}[hb]
	\includegraphics{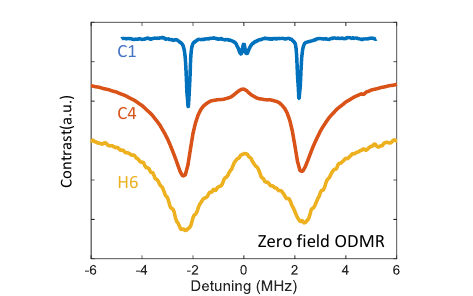}
	\caption{Zero field ODMR measurements on three different samples.}
	\label{Fig_ZeroODMR}
    \end{figure}

    This section presents the growth conditions, post-treatment procedures, and typical measurement data for the samples listed in Table \ref{tab_measurementsresult}. Detailed growth and treatment parameters are summarized in Table \ref{tab_Growthdetails}. Sample C5 contains \(\textsuperscript{15}\mathrm{N}\) rather than \(\textsuperscript{14}\mathrm{N}\) atoms and lacks nuclear spin levels with \( \text{spin} = 0 \), preventing the observation of true zero-ODMR signals. Consequently, the zero-field ODMR technique cannot differentiate between electric field and strain noise contributions in this sample. Figure \ref{Fig_ZeroODMR} presents representative zero-field ODMR measurements from three samples, selected to demonstrate the progression of increasing electric field noise.
    
    More details about HPHT samples are as follows. H1 and H2 were grown at ~6.1 GPa and ~1430℃ by using a Fe-Al metal solvent. As the 12C-enriched solid carbon sources, CVD powder and 12C-enriched pyrolytic carbon powder fabricated from 99.97\% 12C-enriched methane were used for H1 and H2, respectively.
    H3-H6 were grown at 5.5 GPa and ~1350℃ using Fe–Co–Ti solvent and 12C-enriched solid carbon source obtained by pyrolysis of 99.999 \% 12C-enriched methane. The isotopic abundance ratio of the grown crystals was determined by SIMS measurement. Nitrogen concentration was controlled by changing the amount of nitrogen getter.    
    H1 was irradiated at 450 \si{\celsius} by irradiating the crystal on a platinum plate in a sealed silica-glass tube. After two cycles of irradiation, H1 was annealed at 1000 \si{\celsius} for 2 h under vacuum. Samples H2, H3, H4, H5, and H6 were irradiated with 2 MeV electrons at room temperature. Except for H2, the samples were irradiated in multiple steps and annealed (1000\si{\celsius} for 2 h under vacuum) after each irradiation, as shown in the table. \ref{tab_Growthdetails}.

    \begin{table}[ht]
	\caption{Sample Characteristics and Treatment Details}
	\label{tab_Growthdetails}
		\begin{ruledtabular}
			\begin{tabular}{cccccc}
				No. & [$^{12}$C] (\%) & Electrons Irradiation & Annealing    & Initial Nitrogen \\
				\hline
    		  \addlinespace
				H1  & 99.970   &2 MeV,\num{2e17}e/cm\textsuperscript{2}$\times$2 & 1000\si{\celsius} 2 h &0.74 ppm\\
				\addlinespace
				H2  & 99.970   & 2 MeV,\num{7e17} e/cm\textsuperscript{2} & 1000\si{\celsius} 2 h &1.3 ppm \\
				\addlinespace
				H3  & 99.995   &2 MeV,\num{1e17}e/cm\textsuperscript{2}$\times$2 & 1000\si{\celsius} 2 h$\times$2 &3.5 ppm \\
				\addlinespace
				\multirow{2}{*}{H4}
                & \multirow{2}{*}{99.995}
                & 2 MeV,\num{1.5e17} e/cm\textsuperscript{2}
                & \multirow{2}{*}{1000\si{\celsius} 2 h$\times2$}
                & \multirow{2}{*}{1.7 ppm}\\
                \addlinespace
                & &+ 2 MeV,\num{0.5e17} e/cm\textsuperscript{2}
                \\
				\addlinespace
				H5  & 99.995   &2 MeV,\num{1e17}e/cm\textsuperscript{2}$\times$3 & 1000\si{\celsius} 2 h$\times$3 &6 ppm \\
				\addlinespace
				H6  & 99.995   & 2 MeV,\num{1e17}e/cm\textsuperscript{2}$\times$5 & 1000\si{\celsius} 2 h$\times$5 &33 ppm \\
    		    \addlinespace
                \hline
                \addlinespace
				C1  & 99.999   & 1 MeV,\num{4.8e17} e/cm\textsuperscript{2}
                & \multirow{5}{*}{800\si{\celsius} 10 h}
                &0.96 ppm\\
     		\addlinespace   
				C2  & 99.999   & 1 MeV,\num{3.4e17} e/cm\textsuperscript{2}
                &\multirow{5}{*}{+1200\si{\celsius} 1 h}
                &1.9 ppm\\
        		\addlinespace
				C3  & 99.999   & 1 MeV,\num{6.9e17} e/cm\textsuperscript{2}&&3.0 ppm\\
        		\addlinespace    
				C4  & 99.999   & 4.5 MeV,\num{5.8e18} e/cm\textsuperscript{2}&&12.8 ppm\\
        		\addlinespace    
				C5  & 99.999   & 6 MeV,\num{6.0e18} e/cm\textsuperscript{2}&&14 ppm\footnote{Note: $^{15}$N is used in the growth of sample C5, while the other samples are grown with $^{14}$N.}\\
			\end{tabular}
		\end{ruledtabular}
	\end{table}

\begin{figure}[ht]
	\includegraphics{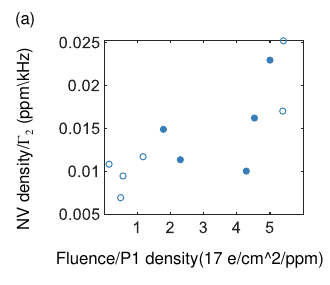}
        \caption{Parameter relation: 1. Ratio between the NV center density and $\Gamma_2$, 2. Ratio between the electron irradiation dose and the P1 center density.}
	\label{Fig_12smple_P1}
    \end{figure}

    Here we analyze the observed linear relationship between NV center concentration and $\Gamma_2$, and explain the deviations observed in certain samples. Previous studies on diamonds with low P1-to-NV conversion rates have identified P1 centers as the primary decoherence source for $T_2$ \cite{4PhysRevB2022Walsworth, RN20, RN18}. These studies established a decoherence contribution factor of 6.3 kHz/ppm for P1 centers \cite{4PhysRevB2022Walsworth}. For cases where NV-NV interactions dominate the decoherence process, the contribution factor is 20.6 kHz/ppm \cite{22RevModPhys2020Walsworth}. In our samples, which exhibit high P1-to-NV conversion efficiencies, $\Gamma_2$ receives contributions from both P1 and NV centers. Moreover, the observed deviations from linearity can be attributed to varying conversion efficiency from P1 to Nv centers. Since the conversion efficiency is directly related to electron irradiation, we verified our interpretation by examining two key parameters: the ratio of NV center density to $\Gamma_2$, and the ratio of electron irradiation dose to initial P1 center density. Both parameters should be proportional to the conversion efficiency, and indeed, as shown in Figure \ref{Fig_12smple_P1}, they exhibit a linear relationship. The electron irradiation doses and initial P1 center concentrations used in this analysis are provided in Table \ref{tab_Growthdetails}.
    
    \section{\label{Appendix F}Extra details of the setup}
    The experimental measurements were performed using a custom-built confocal microscopy setup. The optical excitation system consists of a high-power 532 nm laser (Lighthouse) for NV center excitation, while the fluorescence detection system employs an avalanche photodetector (Thorlabs APD130A2M) coupled with a high-speed data acquisition card (Spectrum M4i.4421) for signal recording. The microwave (MW) driving field was generated using a combination of RF sources: an Apsin Anapico 4010 serving as the local oscillator, mixed with signals from an arbitrary wave generator (Spectrum DN2.6631), and an additional RF source (Apsin Anapico 3000) specifically for DQ and Strain sequence generation. The precise timing control of laser excitation, fluorescence detection, and RF driving was achieved using a pulse generator (Spincore PulseBlasterESR-PRO).

\nocite{*}

\bibliography{Ref}

\clearpage

\end{document}